\documentclass[twocolumn]{aastex6}
\RequirePackage{lineno}
\usepackage{amsmath}
\usepackage{amssymb}
\usepackage{amsthm}
\usepackage{natbib,url,bm}
\usepackage{array}
\usepackage{float}
\usepackage{graphicx}
\usepackage{subfigure}
\usepackage{color}

\newcounter{ichi}
\newcounter{ni}
\newcounter{san}
\newcounter{yon}
\def\be{\begin{equation}}
\def\ee{\end{equation}}
\def\ba{\begin{eqnarray}}
\def\ea{\end{eqnarray}}

\slugcomment{}

\shorttitle{Testing the young NS scenario of FRB 121102}
\shortauthors{Kashiyama \& Murase}

\begin{document}

\title{Testing the young neutron star scenario with persistent radio emission \\ associated with FRB 121102}
\author{Kazumi Kashiyama\altaffilmark{1} and Kohta Murase\altaffilmark{2,3,4,5}}
\altaffiltext{1}{Department of Physics, the University of Tokyo, Bunkyo, Tokyo 113-0033, Japan}
\altaffiltext{2}{Department of Physics, The Pennsylvania State University, University Park, PA 16802, USA}
\altaffiltext{3}{Department of Astronomy \& Astrophysics, The Pennsylvania State University, University Park, PA 16802, USA}
\altaffiltext{4}{Center for Particle and Gravitational Astrophysics, The Pennsylvania State University, University Park, PA 16802, USA}
\altaffiltext{5}{Yukawa Institute for Theoretical Physics, Kyoto University, Kyoto 606-8502, Japan}


\begin{abstract}
Recently a repeating fast radio burst (FRB) 121102 has been confirmed to be an extragalactic event and a persistent radio counterpart has been identified. 
While other possibilities are not ruled out, the emission properties are broadly consistent with Murase et al. (2016) 
that theoretically proposed quasi-steady radio emission as a counterpart of both FRBs and pulsar-driven supernovae. 
Here we constrain the model parameters of such a young neutron star scenario for FRB 121102. 
If the associated supernova has a conventional ejecta mass of $M_{\rm ej}\gtrsim{\rm a \ few}\ M_\odot$, 
a neutron star with an age of $t_{\rm age} \sim 10-100 \ \rm yrs$, an initial spin period of $P_{i} \lesssim$ a few ms, 
and a dipole magnetic field of $B_{\rm dip} \lesssim {\rm a \ few} \times 10^{13} \ \rm G$ can be compatible with the observations. 
However, in this case, the magnetically-powered scenario may be favored as an FRB energy source because of the efficiency problem in the rotation-powered scenario.  On the other hand, if the associated supernova is an ultra-stripped one or the neutron star is born by the accretion-induced collapse with $M_{\rm ej} \sim 0.1 \ M_\odot$, a younger neutron star with $t_{\rm age} \sim 1-10$ yrs can be the persistent radio source and might produce FRBs with the spin-down power. 
These possibilities can be distinguished by the decline rate of the quasi-steady radio counterpart. 
\end{abstract}


\section{Introduction}
Fast radio burst (FRB) 121102 is the only repeating FRB so far identified~\citep{Spitler+14,Spitler+16}. 
\cite{Spitler+16} reported 10 FRBs in a $\sim 3 \ \rm h$ observation. 
The dispersion measure (DM) is $\sim 560 \ \rm cm^{-3} \ pc$ consistent among all the observed FRBs 
while the spectral shape changes drastically burst by burst; the spectral index ranges from $\sim -10$ to $\sim 8$.  Although the repeating FRB may have a different origin than other non-repeating events and the origin and radio emission mechanism of FRBs is still enigmatic, its repeating properties have provided us with profound implications for our understandings of FRBs. 

Recently \cite{Chatterjee_et_al_17} and \cite{Marcote_et_al_17} reported additional FRBs from a position consistent with FRB 121102 
with angular resolution of $\sim$ second and millisecond using the Karl G. Jansky Very Large Array (VLA) and the European VLBI Network (EVN), respectively. 
\cite{Tendulkar_et_al_17} identified their host galaxy as a dwarf star-forming galaxy with a stellar mass of $M_\ast \sim (4-7) \times 10^7 \ M_\odot$ at redshift $z = 0.19273(8)$.
This is the first, solid confirmation that FRBs are cosmological events. 
Furthermore, they identified a quasi-steady radio source in the error circle of the FRB arrival direction. 
The observed flux is $\sim 200 \ \rm \mu Jy$ at $\sim 1-10 \ \rm GHz$, corresponding to a luminosity of a few $\times 10^{39} \ \rm erg \ s^{-1}$. 
The light curve shows a $\sim 10 \ \%$ fluctuation with a time scale of $\sim$ a day, which seems consistent with interstellar scintillation. 
The size of the persistent radio source is constrained to be $\lesssim 0.7 \ \rm pc$ and the position is $\sim$ kpc offset from the center of the host galaxy. 
Since the positioning uncertainty of the FRBs is $\sim 40 \ \rm pc$, the FRBs and persistent radio emission may or may not have the same source at this stage. 
In either case, to understand the physical origin of the persistent emission will be one of the keys to understand the physical origin of this repeating FRB. 

Galactic pulsar wind nebulae (PWNe) are known to be powerful accelerators of electrons and positrons~\citep[e.g.,][]{2006ARA&A..44...17G}, 
and pulsar-aided explosions have also been discussed for many years~\citep[e.g.,][]{1971ApJ...164L..95O,Chevalier_87}.
Interestingly, before the discovery of the host galaxy, \cite{Murase_et_al_16} proposed quasi-radio emission as counterparts of both FRBs and pulsar-driven supernova remnants, if FRBs are powered by a young neutron star (NS). By solving kinetic equations, they calculated time-dependent fluxes $\sim10-100$~yr after the supernova explosion.  The quasi-steady counterpart of FRB 121102 is broadly consistent with the previous theoretical predictions, in which the authors considered young NSs including a magnetar and rotating neutron star as the source of FRBs and calculated associated nebula emission of magnetar remnants and pulsar-driven supernovae including super-luminous supernovae (SLSNe). 
In this work, we provide model-independent constraints on the young NS scenario for FRB 121102.  
Throughout this work, we use the notation $Q = 10^{x} Q_{x}$ in CGS units.

\section{Constraining Young Neutron Star Models}

\subsection{Energy Source of FRBs}
Let us first discuss the energy source of repeating FRBs in the context of the young NS scenario. 
The FRB models can be divided into two categories; magnetically-powered scenario or rotation-powered scenario. 

In the former possibility, including the magnetar model~\citep[e.g.,][]{Popov&Postnov10,Lyuvarsky14}, the energy source is the strong magnetic field in the magnetosphere. 
The source field can be transiently transferred from inside the NS.
The magnetic free energy is released via processes such as reconnection with a light crossing timescale $\sim O({\rm ms})$, possibly resulting in FRBs. 
The intrinsic energy budget can be estimated as 
\begin{equation}\label{eq:E_B}
{\cal E}_{B} \approx B_{\rm *}{}^2 R_{\rm *}{}^3/6 \sim 3 \times 10^{43} \ {\rm erg} \ B_{\rm sur, 13}{}^2,
\end{equation}
where $B_{\rm *}$ is the magnetic field and $R_{\rm *} = 12 \ \rm km$ is the radius of the NS. 
The minimum required energy to explain a repeating FRB is 
\begin{eqnarray}\label{eq:E_FRB_min}
{\cal E}_{\rm FRB, min} &\approx& f_{\rm b} E_{\rm FRB} {\cal R}_{\rm FRB} t_{\rm age}  \notag \\
&\sim& 3 \times 10^{43} \ {\rm erg} \ f_{\rm b} E_{\rm FRB, 39} {\cal R}_{\rm FRB, -5}  t_{\rm age, 9.5},
\end{eqnarray}
where $E_{\rm FRB} \sim 10^{39-40} \ \rm erg$ is the fluence per one FRB, $f_{\rm b}$ is the beaming factor, ${\cal R}_{\rm FRB}$ is the FRB repeat rate, and $t_{\rm age}$ is the age of the NS.  
From Eqs. (\ref{eq:E_B}) and (\ref{eq:E_FRB_min}), the NS magnetic field has to be larger than a threshold value;
\begin{equation}\label{eq:B_sur}
B_{\rm *} \gtrsim 10^{13} \ {\rm G} \ f_{\rm b}{}^{1/2} E_{\rm FRB, 39}{}^{1/2} {\cal R}_{\rm FRB, -5}{}^{1/2} t_{\rm age, 9.5}{}^{1/2} .
\end{equation}
The magnetar model can be motivated by millisecond gamma-ray pulses observed from magnetars including soft-gamma-ray repeaters or anomalous X-ray pulsars. 
However, there has been no simultaneous detection of a magnetar burst and an FRB which can constrain the scenario~\citep{Tendulkar_et_al_16,Murase_et_al_16,Yamasaki_et_al_16,DeLaunay+16frb131104}. 

In the rotation-powered model~\citep[e.g.,][]{Connor+16,Cordes_et_al_16,Lyutikov_et_al_16}, FRBs are considered to be a scaled-up version of Crab giant pulses. 
The intrinsic energy budget is the rotation energy of the NS; 
\begin{equation}
{\cal E}_{\rm rot, i} \approx \frac{1}{2} I (2\pi/P_{i})^2 \sim 1.9 \times 10^{52} \ {\rm erg \ s^{-1}} P_{i, -3}{}^{-2}. 
\end{equation}
Here $I \approx 0.35\times M_{\rm *} R_{\rm *}{}^2$ is the momentum of inertia and $P_{i}$ is the initial spin period and $M_{\rm *} = 1.4 \ M_\odot$ is the NS mass. 
The rotation energy can be extracted by the unipolar induction. 
The initial spin-down luminosity is 
\begin{eqnarray}
L_{\rm sd, i} &\approx& \frac{B_{\rm dip}^2 (2\pi/P_{i})^4 R^6}{4c^3} (1+\sin^2\chi) \notag \\
&\sim& 2.4 \times 10^{44} \ {\rm erg \ s^{-1}} B_{\rm dip, 12.5}{}^2 P_{i, -3}{}^4, 
\end{eqnarray}
where $B_{\rm dip}$ is the dipole magnetic field strength and $\chi$ is the angle between the axes of rotation and dipole magnetic field~\citep{Gruzinov_05,Spitkovsky_06,Tchekhovskoy_et_al_13}. 
Hereafter we assume that $B_{\rm dip}$ and $\sin^2 \chi = 2/3$ are time-independent. 
The spin period and spin-down luminosity are almost constant $P \approx P_{i}$, $L_{\rm sd} \approx L_{\rm sd, i}$ for $t \lesssim t_{\rm sd}$ where 
\begin{equation}
t_{\rm sd} \approx \frac{{\cal E}_{\rm rot, i}}{L_{\rm sd, i}} \sim 2.5 \ {\rm yrs} \ B_{\rm dip, 12.5}{}^{-2} P_{i, -3}{}^{2}
\end{equation}
while they evolve as $ P \approx P_{i} (t/t_{\rm sd})$, $L_{\rm sd} \approx L_{\rm sd, i} \times (t/t_{\rm sd})^{-2}$ for $t \gtrsim t_{\rm sd}$. 
In order to power an FRB, the spin-down luminosity of fast-spinning young NSs should be larger than the FRB luminosity; $L_{\rm sd, i} \times (t_{\rm age}/t_{\rm sd})^{-2} > f_{\rm b} L_{\rm FRB}$ or 
\begin{equation}\label{eq:L_sd_cond}
t_{\rm age} \lesssim 39 \ {\rm yrs} \ f_{\rm b}^{-1/2} L_{\rm FRB, 42}{}^{1/2} B_{\rm dip, 12.5}{}^{-2} P_{i, -3}{}^{2}.
\end{equation}
Here $L_{\rm FRB} \sim 10^{42-43} \ \rm erg \ s^{-1}$ is the characteristic isotropic luminosity of FRBs. 

Eqs (\ref{eq:B_sur}) and (\ref{eq:L_sd_cond}) are the energy condition for the magnetar and rotation-powered scenario, respectively. 
Although $f_{\rm b}$ and ${\cal R}_{\rm FRB}$ are largely uncertain at this stage, a younger NS can more easily satisfy these conditions for a fixed $(B_{\rm dip}, B_{\rm *}, P_{i})$.

\subsection{Evolution of Pulsar Wind Nebulae and Supernova Remnants}
In general, young NSs should be surrounded by a PWN and supernova remnant (SNR), 
both of which play important roles to determine observational properties of FRBs~\citep[][for discussions]{Murase_et_al_16}. 
In particular, both the PWN and SNR can scatter or absorb FRBs and also provide a large dispersion which may conflict with the observed DM of FRB 121102
\citep{Kulkarni_et_al_14,Murase_et_al_16,Piro_16,Lyutikov_et_al_16}.   
These basically set the lower limit for the age of the NS. 
On the other hand, in order to explain the observed quasi-steady radio counterpart by the PWN emission, the NS should be energetic and young enough. 
By combining these conditions with Eqs. (\ref{eq:B_sur}) and (\ref{eq:L_sd_cond}), the parameter space of the young NS model can be constrained. 
To this end, we have to solve consistently the spin-down of the central NS and the (magnetohydro)dynamical evolution of the PWN and SNR including the ionization state in the SN ejecta.
The PWN emission consists of the superposition of synchrotron and inverse Compton emission from relativistic electrons. Although the injection and acceleration mechanisms are still uncertain, 
it is natural to apply the phenomenological results on Galactic PWNe~\citep[e.g.,][and references therein]{Gaensler_et_al_06,Tanaka_Takahara_10} to extragalactic PWNe with $t_{\rm age}\gtrsim$~a few~yr.  

We here calculate the time evolution of the spin-down luminosity $L_{\rm sd}(t)$ and the radii of the PWN and SN ejecta $r_{\rm nb}(r)$, $r_{\rm ej}(t)$, using a simplified model shown in Appendix of \cite{Kashiyama_et_al_16} \citep[see also][]{1992ApJ...395..540C,2005ApJ...619..839C,2015ApJ...805...82M}.  
Other relevant physical quantities can be calculated from them with adding some phenomenological parameters. 
For example, we estimate the mean magnetic field in the PWN as 
\begin{equation}
B_{\rm nb} \approx \left[ \frac{6 \epsilon_{\rm B}}{r_{\rm nb}(t){}^3} \int^{t_{\rm age}} L_{\rm sd}(t') dt' \right]^{1/2},
\end{equation}
where $\epsilon_{\rm B}$ is the magnetic-field amplification efficiency, which we fix as $\epsilon_{B} = 0.01$. 
Also, the number density of free electrons in the SN ejecta is given by $n_{\rm e, ej} \approx 3M_{\rm ej}/(4\pi r_{\rm ej}{}^3 \mu_{\rm e} \bar A m_{\rm H})$, where $\bar A$ is the mean atomic mass number, which we set $\bar A = 10$.  As for the ionization state, we assume the singly ionized state ($\mu_{\rm e} \approx 1$) and correspondingly set the electron temperature as ${\cal T}_{\rm e} \sim 10^4 \ \rm K$. 
Rigorous calculation of the PWN emission is also beyond the scope of this paper. 
Instead we estimate the minimum required energy for reproducing the observed quasi-steady radio counterpart 
and check whether each NS model can provide a sufficient energy. 

\subsection{Physical Conditions}
Now we show the conditions imposed on the young NS model obtained from the observations of FRB 121102.  

First, FRBs produced in the NS magnetosphere or possibly in the PWN can be degraded by scattering and absorption processes in the PWN and SNR. 
The most relevant process is the free-free absorption in the SN ejecta~\citep[e.g.,][]{2016arXiv161103848M}.
The NS needs to be old enough so that the opacity at $\sim \rm GHz$ frequencies becomes small enough; 
\begin{equation}\label{eq:tau_ff}
\tau_{\rm ff} |_{\nu = 1 \rm GHz} \approx 2.1 \times 10^{-25} \ {\cal T}_{\rm e,4}^{-1.35} \int dr \ n_{\rm e, ej} n_{\rm i, ej} \bar Z^2< 1,   
\end{equation}
where $\bar Z \sim \bar A/2$. 

Second, propagation of an FRB through the SN ejecta can induce a significant amount of DM. 
In the case of FRB 121102, DM from the host galaxy and near source region is estimated to be $55 \ {\rm pc \ cm^{-3}} \lesssim \rm DM_{\rm host} \lesssim 225 \ {\rm pc \ cm^{-3}}$. 
An interstellar propagation through the host galaxy can naturally provide $O(100) \ \rm pc \ cm^{-3}$~\citep{Tendulkar_et_al_17}. 
The observed DM of FRB 121102 is roughly constant during the observed period $\sim 4 \ \rm yrs$ 
while the DM due to the SNR should evolve as $\rm DM_{\rm ej} \propto t^{-2}$~\citep{Murase_et_al_16,Lyutikov_et_al_16}. 
The NS also needs to be old enough so that 
\begin{equation}\label{eq:dm}
{\rm DM}_{\rm ej} \approx n_{\rm e, ej} r_{\rm ej} \ll 100 \ \rm pc \ cm^{-3}
\end{equation}
is satisfied. 

Third, the observed persistent radio emission of FRB 121102 can be interpreted as synchrotron emission from relativistic electrons in the PWN.  
Given that the power budget of the PWN is the spin-down luminosity, the energy condition can be described as 
\begin{equation}\label{eq:E_e,nb}
{\cal E}_{\rm e, nb, min} < {\cal E}_{\rm e, inj, max},
\end{equation}
where $\cal E_{\rm e, nb, min}$ is the required energy in relativistic electrons in order to explain the quasi-steady radio emission 
and $\cal E_{\rm e, inj, max}$ is the upper limit of the energy stored in the PWN. 
We estimate $\cal E_{\rm e, inj, max}$ as 
\begin{eqnarray}\label{eq:E_inj,max}
{\cal E}_{\rm e, inj, max}  \approx && \epsilon_{\rm e} \int^{t_{\rm age}} L_{\rm sd}(t') dt' \notag \\ 
&\times& \begin{cases}
    1 & (t_{\rm age} < t_{\rm sd}) \\
    \frac{t_{\rm sd}}{t_{\rm age}} [1+\log(\frac{t_{\rm age}}{t_{\rm sd}})] & (t_{\rm age} > t_{\rm sd})
  \end{cases},
\end{eqnarray}
where $\epsilon_{\rm e} \approx 1- \epsilon_B$ is the injection efficiency. 
The last factor in the right hand side represents the effect of adiabatic energy loss of the PWN.
In fact, the radiative energy loss can be also relevant and thus  Eq. (\ref{eq:E_inj,max}) gives a strict upper limit. 
The observed quasi-steady radio spectrum of FRB121102 can be fitted by a broken power law~\citep{Chatterjee_et_al_17}; 
\begin{equation}
  \nu F_\nu \approx \frac{{\cal L}_{\rm nb}}{4\pi d_L^2} \times \begin{cases}
    (\nu/\bar \nu)^{p_1} & (\nu < \bar \nu) \\
    (\nu/\bar \nu)^{p_2} & (\nu > \bar \nu) 
  \end{cases},
\end{equation}
with ${\bar \nu} \sim 10 \ \rm GHz$, ${\cal L}_{\rm nb} \sim 1.9 \times 10^{39} \ \rm erg \ s^{-1}$, $p_1 \sim 0.82$, and $p_2 \sim - 0.17$.
Although the $\nu F_\nu$ spectrum may have other peaks, we here assume the PWN bolometric luminosity is dominated by the radio bands and estimate the minimum required energy stored in PWN electrons.  
The typical Lorentz factor of radio-emitting electrons can be calculated as ${\bar \gamma_{\rm e}} \approx (4\pi m_{\rm e} c \bar \nu/3e B_{\rm nb})^{1/2}$. 
The energy loss rate is ${\bar P_{\rm e}} \approx (4/3) \times \sigma_{\rm T} c {\bar \gamma_{\rm e}}{}^2 B_{\rm nb}{}^2/8 \pi$. 
The total number of such electrons in the PWN should be $\bar N_{\rm e} \approx {\cal L}_{\rm nb}/{\bar P_{\rm e}}$.
and then the minimum required energy in the PWN is ${\cal E}_{\rm e, nb, min} \approx \bar N_{\rm e} \times {\bar \gamma_{\rm e}} m_{\rm e} c^2$, or 
\begin{equation}
{\cal E}_{\rm e, nb, min} \approx \frac{3\sqrt{3} (\pi e m_{\rm e}c)^{1/2} {\cal L}_{\rm nb}}{\sigma_{\rm T} {\bar \nu}^{1/2} B_{\rm nb}{}^{3/2}} . 
\end{equation}
In order to satisfy Eq. (\ref{eq:E_e,nb}), the initial spin-down luminosity should be large enough for providing sufficient energy to the PWN 
and the NS should be young enough for preventing significant adiabatic cooling.  

Another important constraint on the PWN may come from the fact that there is no sign of synchrotron self absorption in the persistent radio spectrum~\citep{Yang_et_al_16,Murase_et_al_16}; 
\begin{equation}\label{eq:tau_sa}
\tau_{\rm sa} |_{\nu = 1 \rm  GHz} < 1.
\end{equation}
In order to estimate the self-absorption optical depth, one has to specify the energy spectrum of synchrotron emitting electrons~\citep[e.g.,][]{Rybicki_Lightman}. 
Here we consider the minimum set of electrons, which can reproduce the persistent radio emission; 
\begin{equation}
  \frac{dn_{\rm e, nb}}{d\gamma_{\rm e}} \approx \frac{3{\bar N}_{\rm e}}{4\pi r_{\rm nb}{}^3{\bar \gamma_{\rm e}} } \begin{cases}
    (\gamma_{\rm e}/{\bar \gamma_{\rm e}})^{-q_1} & (\gamma_{\rm e} < {\bar \gamma_{\rm e}}) \\
    (\gamma_{\rm e}/{\bar \gamma_{\rm e}})^{-q_2}& ({\bar \gamma_{\rm e}} < \gamma_{\rm e})
  \end{cases},
\end{equation}
with $q_1 \approx 3-2p_1 \approx 1.4$, $q_2 \approx 3-2p_2 \approx 2.7$.
Note that the electrons with $\gamma_{\rm e} < {\bar \gamma_{\rm e}}$ are basically in the slow cooling regime for the cases we are interested in. 
This gives a conservative estimate of $\tau_{\rm sa}$, thus resulting constraints on the model parameters are also conservative. 
Assuming the electron spectrum above, one can also estimate the minimum DM contribution from the PWN, which should be also much smaller than the host galaxy contribution; 
\begin{equation}\label{eq:dm_nb}
{\rm DM}_{\rm nb} \approx n_{\rm e, nb} r_{\rm nb} \approx \frac{3{\bar N}_{\rm e}{\bar \gamma_{\rm e}}{}^{q_1 -1}}{4\pi r_{\rm nb}{}^2} \ll 100 \ \rm pc \ cm^{-3}
\end{equation}

Lastly and importantly, the size of the PWN should be smaller than the observed upper limit on the size of the persistent radio source~\citep{Marcote_et_al_17}; 
\begin{equation}\label{eq:r_nb}
r_{\rm nb} < 0.7 \ \rm pc. 
\end{equation}

\subsection{Constraints on the Model Parameters}
\begin{figure}
\begin{center}
\includegraphics[width=0.8\columnwidth]{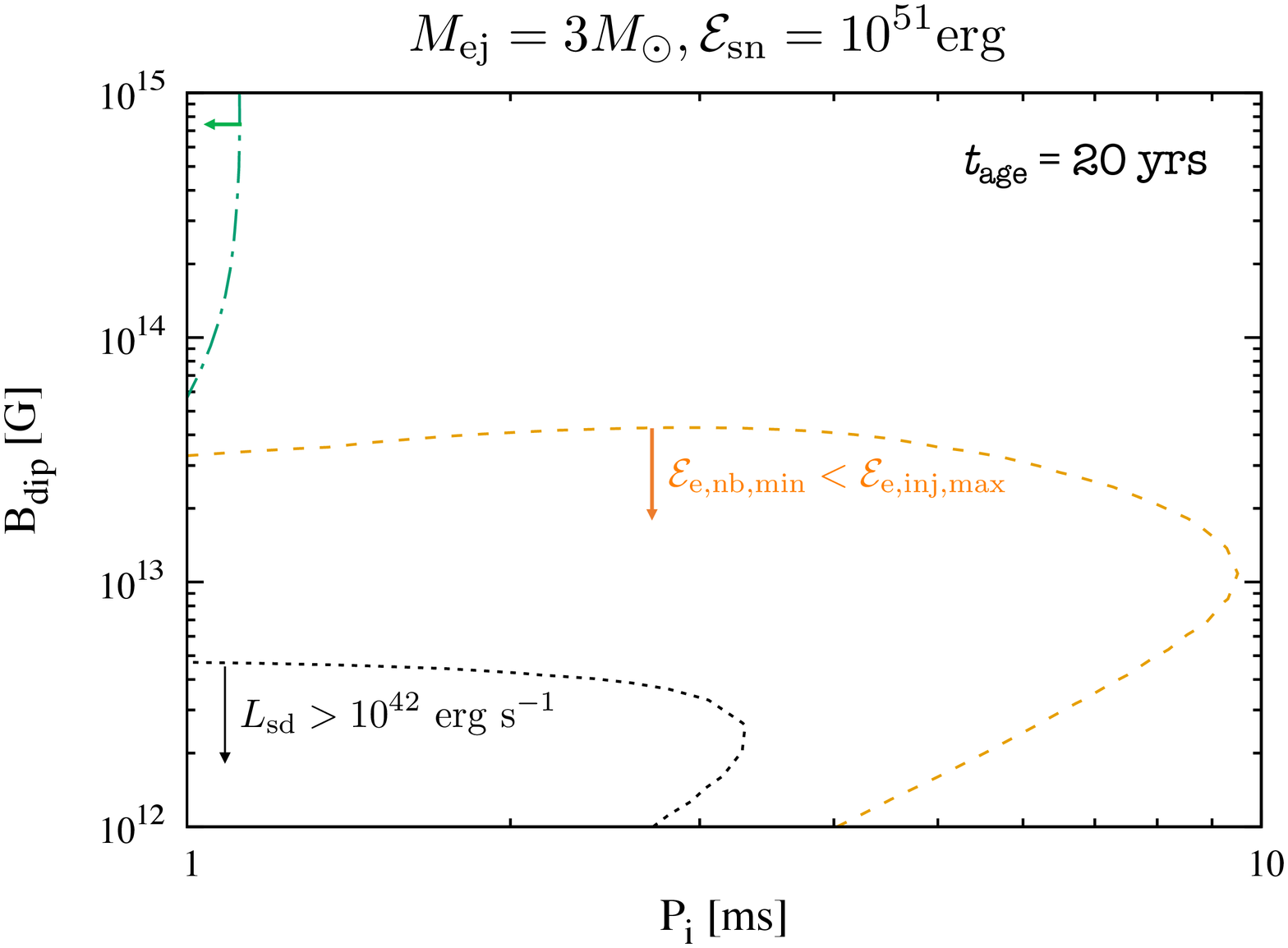}
\includegraphics[width=0.8\columnwidth]{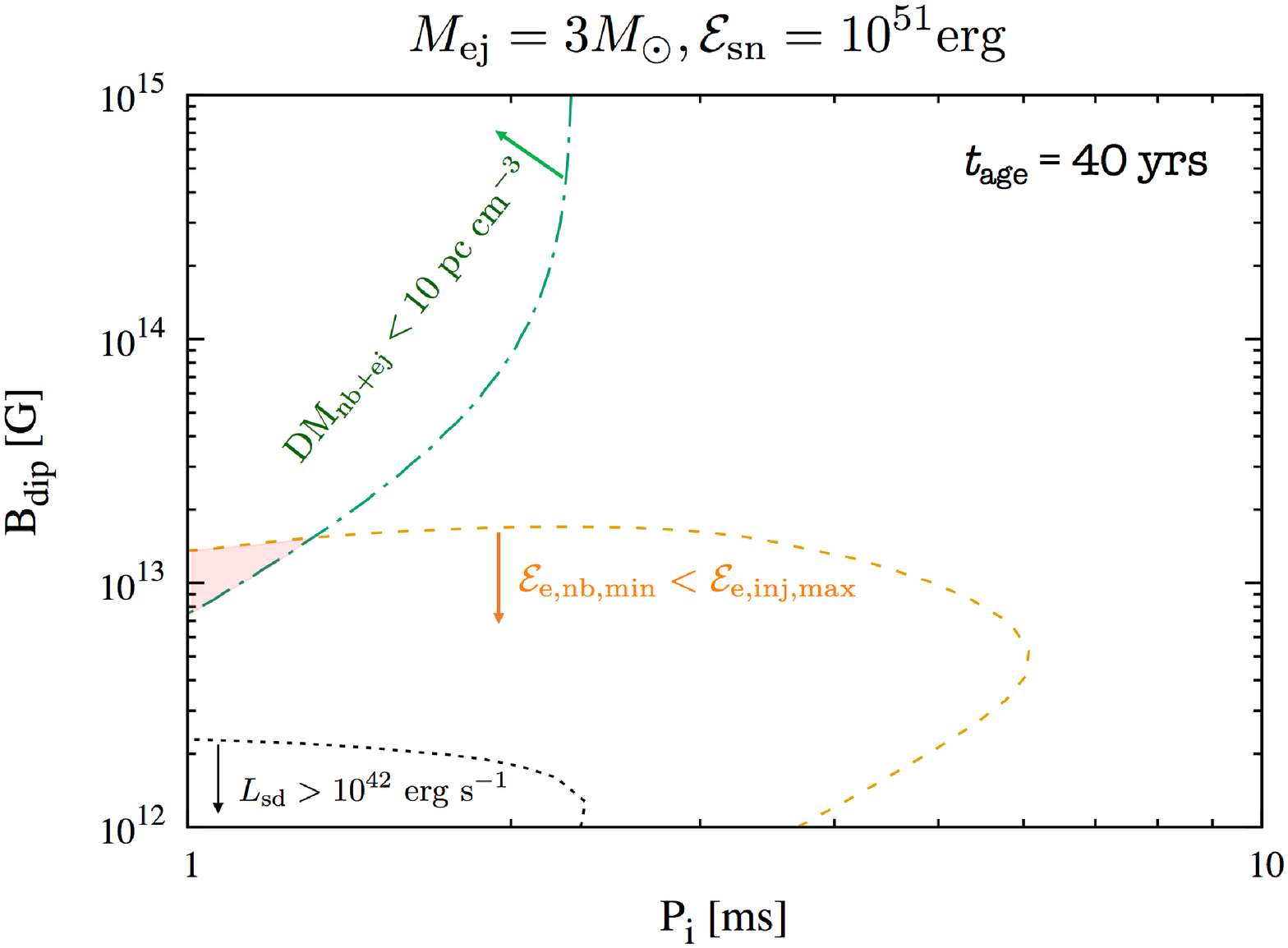}
\includegraphics[width=0.8\columnwidth]{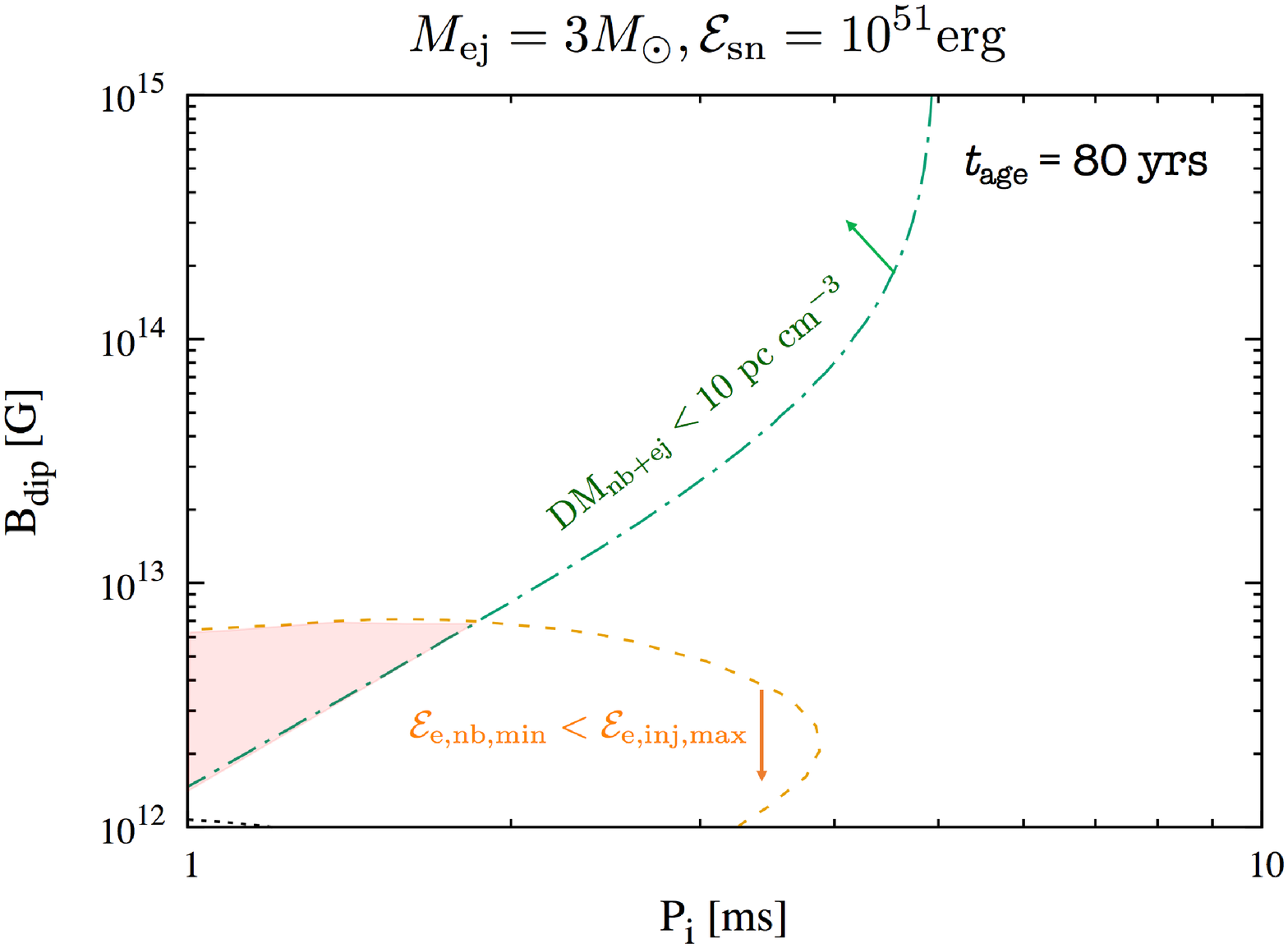}
\caption{
Constraints on the parameter space $(P_{i}, B_{\rm dip})$ of the young NS model for FRB 121102. 
We set the SN ejecta mass and explosion energy as ($M_{\rm ej}, {\cal E}_{\rm sn}$) = ($3 \ M_\odot, 10^{51} \ \rm erg$).
The dashed and dotted-dash lines indicate the minimum energy requirement of the PWN (Eq. \ref{eq:E_e,nb}) 
and the condition on the DM contribution from the PWN and SN ejecta (Eqs. \ref{eq:dm} and \ref{eq:dm_nb}), respectively.
The SN ejecta is assumed to be singly ionized, for simplicity.
The dotted lines enclose the parameter space in which the spin-down luminosity can be comparable to the FRB luminosity (Eq. \ref{eq:L_sd_cond}).
}\label{fig:Ibc}
\end{center}
\end{figure}

\begin{figure}
\begin{center}
\includegraphics[width=0.8\columnwidth]{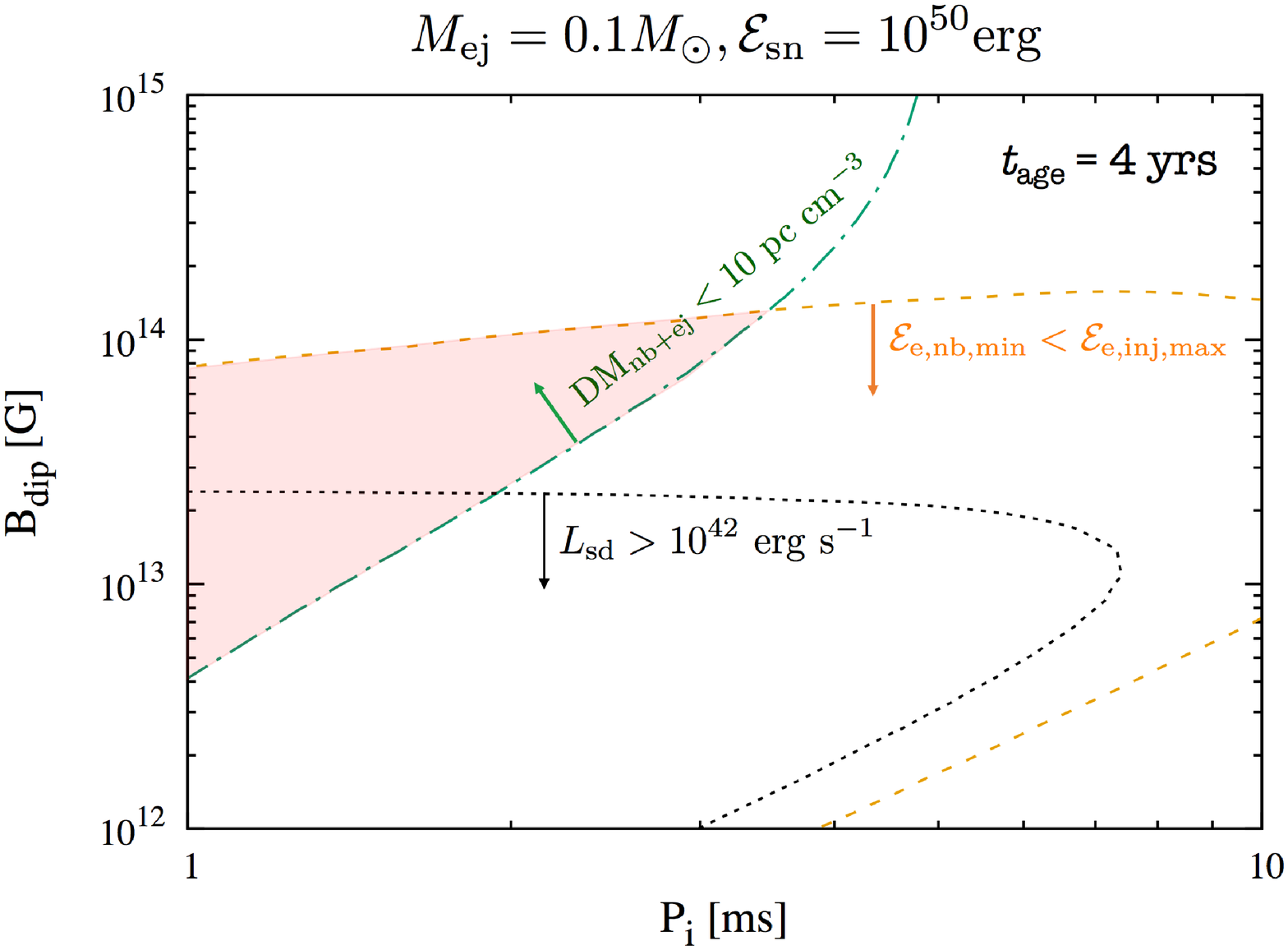}
\includegraphics[width=0.8\columnwidth]{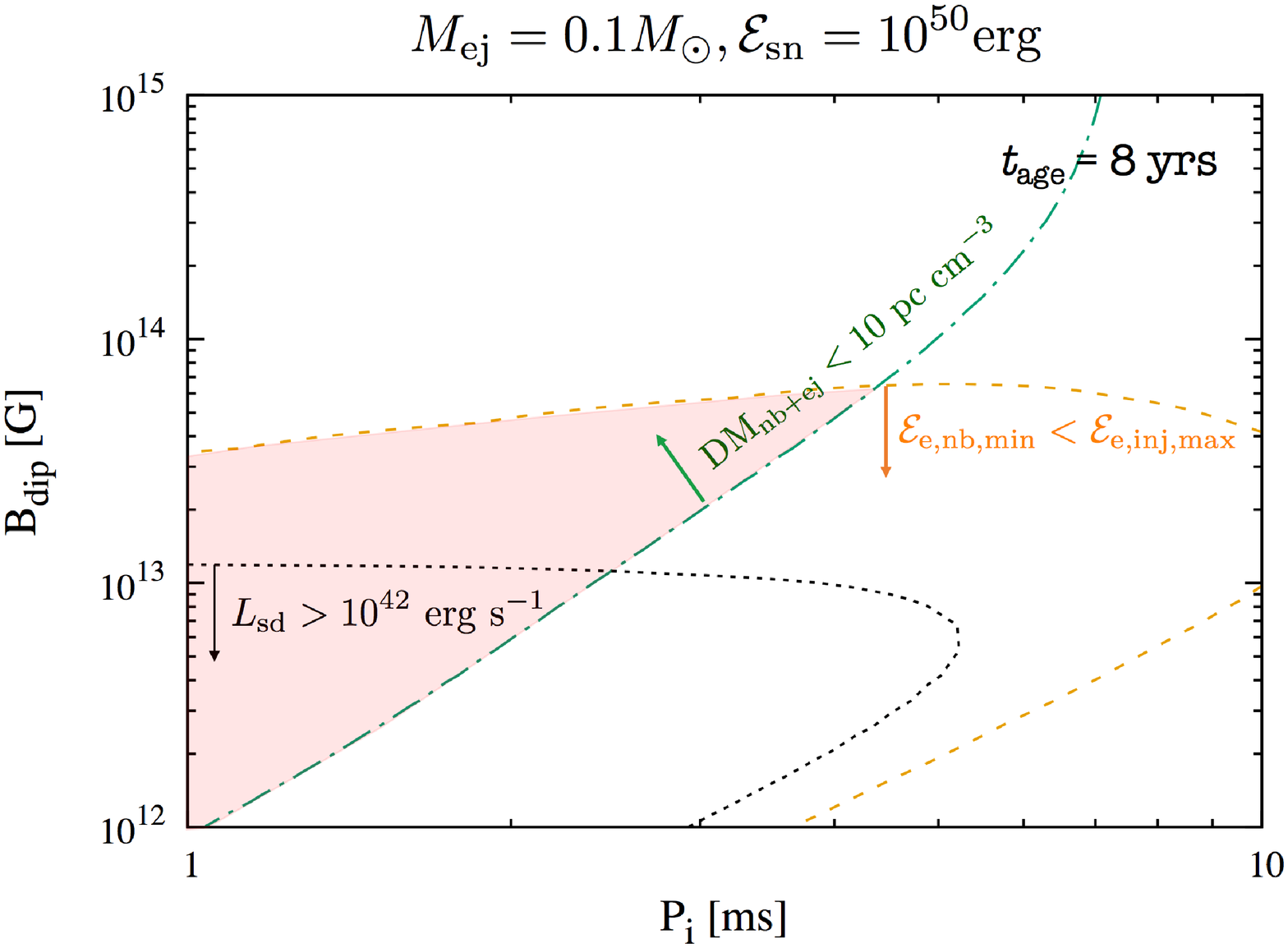}
\includegraphics[width=0.8\columnwidth]{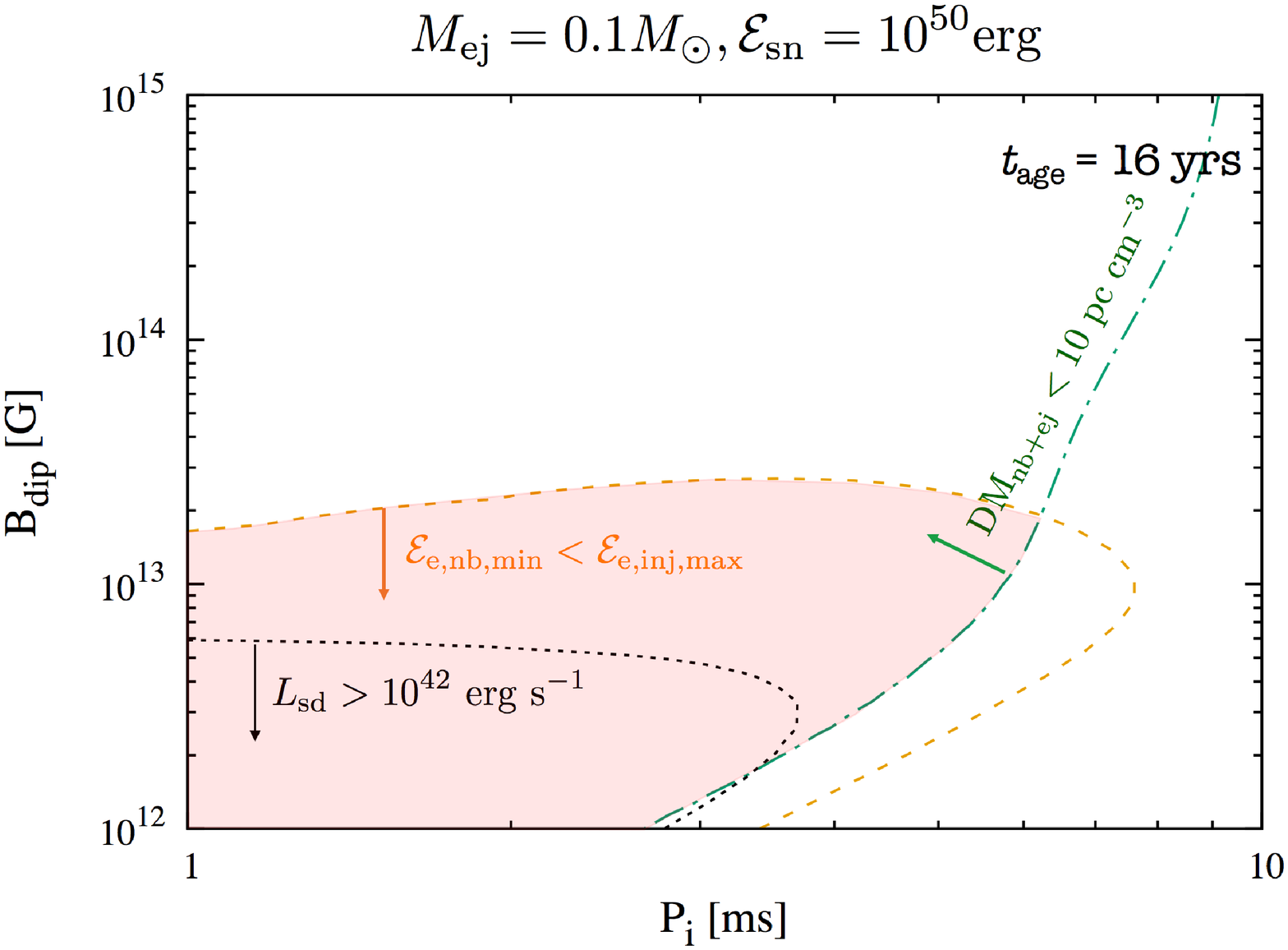}
\caption{
Same as Fig. \ref{fig:Ibc} but with ($M_{\rm ej}, {\cal E}_{\rm sn}$) = ($0.1 \ M_\odot, 10^{50} \ \rm erg$), corresponding to ultra-stripped SNe, 
and $t_{\rm age} =$ 4, 8, and 16 yrs from the top to the bottom. 
}\label{fig:US}
\end{center}
\end{figure}

Now let us discuss with what parameters the young NS model can satisfy all the necessary conditions (Eqs. \ref{eq:tau_ff}, \ref{eq:dm}, \ref{eq:E_e,nb}, \ref{eq:tau_sa}, \ref{eq:dm_nb}, and \ref{eq:r_nb}).
The input parameters are the initial spin period $P_{i}$, the dipole magnetic field $B_{\rm dip}$, the SN ejecta mass $M_{\rm ej}$, and the SN explosion energy $\cal E_{\rm sn}$.  
Note that $\cal E_{\rm sn}$ determines to the initial ejecta velocity while the ejecta is accelerated by the pulsar wind.

Fig. \ref{fig:Ibc} shows the constraints on the parameter space of $(P_{i}, B_{\rm dip})$ for the case with ($M_{\rm ej}, {\cal E}_{\rm sn}$) = ($3 \ M_\odot, 10^{51} \ \rm erg$), 
corresponding to conventional core-collapse SNe. 
The top, middle, and bottom panels shows the cases with $t_{\rm age} =$ 20, 40, and 80 yrs, respectively.  
In this case, it turns out that the condition on the DM contributions from the PWN and SN ejecta (Eqs. \ref{eq:dm} and \ref{eq:dm_nb}; dotted-dash line) 
and the minimum energy requirement for the PWN (Eq. \ref{eq:E_e,nb}; dashed line) are the most constraining.  
When an NS is very young ($t_{\rm age} \lesssim 30 \ {\rm yrs}$), the PWN and SN ejecta are dense enough so that they can provide a large DM inconsistent with the observation. 
The DM contributions decrease with the NS age and allowed parameter space expands. 
For a given $P_{i}$, ${\rm DM}_{\rm ej+nb}$ decreases faster for a larger $B_{\rm dip}$ since more energies are injected into the ejecta with a shorter timescale and the ejecta expands faster. 
On the other hand, the parameter space satisfying the PWN energy requirement shrinks with time 
and a smaller $B_{\rm dip}$ is preferred for a given $P_{i}$ in order to prevent a significant adiabatic cooling.  
Resultantly, the parameter space satisfying the both conditions is limited (shaded region), $P_{i} \sim 1 \ \rm ms$, $B_{\rm dip} \sim 10^{12-13} \ \rm G$, and $t_{\rm age} \sim 30-100 \ \rm yrs$.  
In Fig. \ref{fig:Ibc}, we also indicate the parameter space satisfying Eq. (\ref{eq:L_sd_cond}) with $f_{\rm b} = 1$ (dotted lines), which does not overlap with the shaded regions. 
This means that the energy source of FRB is most likely the NS magnetic field for the cases with $M_{\rm ej} > {\rm a \ few} \ M_\odot$. 

To show the dependence of the constraints on the SN parameters, we also calculate the cases with ($M_{\rm ej}, {\cal E}_{\rm sn}$) = ($0.1 M_\odot, 10^{50} \ \rm erg$) (Fig. \ref{fig:US}).
This may correspond to the so-called ultra-stripped SNe~\citep[e.g.,][]{Kleiser_Kasen_14,Tauris_et_al_15,Suwa_et_al_15}. 
The progenitors are considered to be in massive close binaries and significantly lose their envelope by the binary interaction. 
Also, accretion-induced collapse (AIC) of white dwarfs may result in a similar explosion~\citep[e.g.,][]{Piro_Kulkarni_13,Moriya_16}.
In Fig. \ref{fig:US}, we show the cases with $t_{\rm age} =$ 4, 8, and 16 yrs from the top to the bottom. 
The PWN energy requirement condition can be more easily satisfied since the NSs are relatively young and energetic 
while the DM contribution from the PWN and SN ejecta can become small enough due to the small ejecta mass. 
As a result, the allowed parameter space becomes larger than the previous case.
Namely, the parameter space satisfying Eq. (\ref{eq:L_sd_cond}) overlaps with the shaded region, 
meaning that the energy source of FRB can be the spin-down luminosity for the cases with $M_{\rm ej} \sim 0.1 \ M_\odot$. 

\section{Discussion}
We have constrained the parameters of the young NS model for FRB 121102 with the quasi-steady radio counterpart.  
The constraints are set by combining the minimum energy requirement for the PWN (Eq. \ref{eq:E_e,nb}) 
and the condition that the DM contribution from the PWN and SN ejecta should be significantly smaller than the observed DM of FRB 121102 (Eqs. \ref{eq:dm} and \ref{eq:dm_nb}). 
Eq. (\ref{eq:E_e,nb}) is conservative in the sense that we only take into account the adiabatic cooling of PWN and assume the minimum set of electrons to explain the quasi-steady counterpart.    
On the other hand, we might overestimate ${\rm DM}_{\rm ej+nb}$ if the neutralization of the free electrons proceeds efficiently in the SN ejecta. 
While there is uncertainty, our results imply that if the associated SN has a conventional ejecta mass $M_{\rm ej} \gtrsim {\rm a \ few} \ M_\odot$, 
NSs with $t_{\rm age} \sim 10-100 \ \rm yrs$, $P_{i} \lesssim$ a few ms, and $B_{\rm dip} \lesssim$ a few $\times 10^{13} \ \rm G$ can be compatible with the observations of FRB 121102.
If the SN ejecta mass is as small as $M_{\rm ej} \sim 0.1 \ M_\odot$, as expected for ultra-stripped SNe or AICs, younger NSs with $t_{\rm age} \sim 1-10 \ \rm yrs$ can be the source. 
The deceleration timescale of the SN ejecta with $M_{\rm ej} \sim 0.1 \ M_\odot$ can be $\sim 10 \ \rm yrs$ and then the PWN may be disrupted by the reverse shock.
Thus, $t_{\rm age} \lesssim 10 \ \rm yrs$ would be preferred for such cases. 

The constraints obtained in this work have useful implications for the energy source of FRBs.
As for the cases with $M_{\rm ej} \gtrsim {\rm a \ few} \ M_\odot$, the spin-down luminosity at $t_{\rm age} \sim 10-100 \ \rm yrs$ already becomes significantly smaller than the observed isotropic luminosity of FRB. 
Thus, the magnetic energy confined in the NS or its magnetosphere is preferred as the energy source of FRBs.
On the other hand, the efficiency problem in the rotation-powered scenario might be overcome in the cases with $M_{\rm ej} \sim 0.1 \ M_\odot$ thanks to the younger ages, 
but either the magnetic field or spin-down power could be the energy source of FRBs. 

The above two possibilities can be distinguished by (near) future observations. 
For example, the age of the NS could be inferred from follow-up observations of the quasi-steady radio counterpart; 
the decline rate is presumably the current spin-down timescale of the NS. 
To this end, more detailed modeling of the PWN emission is necessary. 

The optimal parameter set for the cases with $M_{\rm ej} \gtrsim {\rm a \ few} \ M_\odot$ are marginally consistent with those assumed in the magnetar model for SLSNe~\citep[e.g.,][]{Woosley_10,Kasen_Bildsten_10}\footnote{The optimal values of $B_{\rm dip}$ and $P_{i}$ for explaining SLSNe may change by a factor of a few depending on whether one uses the classical dipole formula or the one obtained by force-free simulations for the NS spin-down luminosity~\citep{Kashiyama_et_al_16}.} 
while bright fast UV-optical transients would associated with the cases with $M_{\rm ej} \sim 0.1 \ M_\odot$.  
Type-I SLSNe preferentially occur in dwarf-star-forming galaxies~\citep{Perley_et_al_16}, which is also consistent with the observed host of FRB 121102~\citep{Chatterjee_et_al_17,Tendulkar_et_al_17}, 
while the typical observed hosts of ultra-stripped SNe look inconsistent with those~\citep{Drout_et_al_14}.
This may hint at a connection between FRB and SLSN progenitors as previously pointed out by \cite{Murase_et_al_16}. 
If this is the case, the formation of (repeating) FRBs may be rare, $\sim 0.01 \ \%$ of core-collapse SNe~\citep[e.g.,][]{Quimby_et_al_13}. 
On the other hand, the event rate of ultra-stripped SNe or AICs is estimated as $\sim 0.1-1 \ \%$ of core-collapse SNe~\citep[e.g.,][]{Tauris_et_al_13,Ruiter_et_al_09}.  
The difference of the event rate could also be the key to discriminate the possibilities when the number density of repeating FRB is well determined. 
 
\section*{Acknowledgements}
The work of KM is supported by NSF Grant No. PHY-1620777.  While we are completing the draft, \cite{2017arXiv170102003L}, \cite{2017arXiv170102370M}, and \cite{2017arXiv170104094Z} appeared. 
Our conclusion on the energy source of FRB 121102 is consistent with \cite{2017arXiv170102003L}, and \cite{2017arXiv170102370M} shares some points with our early work~\citep{Murase_et_al_16} regarding the connection between FRBs and SLSNe.


\end{document}